\documentstyle[12pt]{article}
\begin{document}

\begin{titlepage}

\noindent Stockholm\\
USITP 96-16\\
December 1996

\vspace{2cm}

\begin{center}

{\Large MANIFEST DUALITY IN BORN-INFELD THEORY}

\vspace{2cm}

{\large Ingemar Bengtsson}\footnote{Email address: 
ingemar@vana.physto.se. Supported by NFR.}

\vspace{1cm}

{\sl Fysikum\\
Stockholm University\\
Box 6730, S-113 85 Stockholm, Sweden}

\vspace{15mm}

{\bf Abstract}

\end{center}

\vspace{1cm}

\noindent Born-Infeld theory is formulated using an infinite set 
of gauge fields, along the lines of McClain, Wu and Yu. In this 
formulation electromagnetic duality is generated by a fully local 
functional. The resulting consistency problems are analyzed and 
the formulation is shown to be consistent. 

\end{titlepage}

\noindent {\bf 1. INTRODUCTION.}

\vspace{5mm}

\noindent The present paper has three ingredients: Electromagnetic 
duality, a novel formulation of field theories that uses an infinite 
set of gauge fields to describe a finite number of degrees of freedom 
per spatial point, and Born-Infeld theory. 

Of the many "dualities" now being studied in mathematical 
physics electromagnetic duality stands almost alone in that the 
duality transformation is known explicitly as a function of the 
basic fields of the theory. Even so, it is a non-local transformation 
which is --- as it were --- easily studied only because an abelian 
gauge theory without matter is so simple. 
This is where the second ingredient of the paper enters: If one adopts 
a remarkable suggestion due to McClain, Wu and Yu \cite{McClain} 
electromagnetic duality is in fact generated by a fully local 
functional. The price that one has to pay for this is that an infinite 
set of gauge fields are brought in to describe the two degrees of 
freedom per spatial point that are present in 
the Maxwell theory, and one encounters new consistency problems which 
are akin to those which arise when one makes the transition from 
analytical mechanics to field theory in the first place. These problems 
can be handled however \cite{Kleppe} \cite{Nathan}. Now gauge fields 
were indeed invented to ensure manifest locality of otherwise a
pparently non-local systems. To use an infinite set of gauge fields 
seems a drastic measure, but it may be that 
such measures will prove useful elsewhere --- perhaps to give 
explicit expression to more interesting dualities. Unfortunately, a 
generalization of the McClain-Wu-Yu proposal to Yang-Mills theory 
is not possible without a non-trivial insight as input. This we do 
not have, but it nevertheless seems to be of some interest to study 
the consistency problems of such a formulation for a genuinely 
non-linear theory. This is why we turn to the non-linear theory of 
electrodynamics first studied by Born and Infeld in the thirties 
\cite{Born}. Needless to say the Born-Infeld 
theory is no longer a viable theory of electromagnetism, but it 
arises as a part of an effective action derived from string theory 
as well as in certain worldsheet actions for "branes". It 
also has some remarkable mathematical properties, notably its causal 
behaviour, which sets it apart from other non-linear versions of 
electrodynamics \cite{Plebanski}. For these reasons it retains some 
intrinsic interest. This theory also exhibits electromagnetic duality, 
and as we shall see it admits a formulation along the lines of McClain, 
Wu and Yu. The purpose of this paper is to show that this formulation 
is indeed a consistent one. 

In section 2 we discuss the electromagnetic duality of the Maxwell and 
Born-Infeld theories, in section 3 we discuss the McClain-Wu-Yu 
proposal as applied to linear electrodynamics, and in section 4 we 
give and analyze the corresponding formulation for the Born-Infeld 
theory. About the first two topics we have nothing new to contribute, 
but we spend some time on them because we want to present them in such 
a way that the discussion in section 4 --- which is new --- can be 
kept brief. Our conclusion is stated in section 5.

\vspace{1cm}

\noindent {\bf 2. MAXWELL, BORN-INFELD, AND DUALITY.}

\vspace{5mm}

\noindent The first thing we have to do is to introduce a 
vector potential and a field strength in the usual way. Then 
we form the two Lorentz scalars 

\begin{equation} I_1 = - \frac{1}{4}F_{{\alpha}{\beta}}
F^{{\alpha}{\beta}} \hspace{2cm} I_2 = \frac{1}{4}
F_{{\alpha}{\beta}}\star F^{{\alpha}{\beta}} 
\ , \end{equation}

\noindent where the star denotes the Hodge dual. (We keep spacetime 
flat throughout the paper, since the generalization to curved 
spacetime is straightforward.) If we define electric and magnetic 
fields as

\begin{equation} E_a \equiv F_{ta} \hspace{2cm} B_a \equiv 
\frac{1}{2}{\epsilon}_{abc}F_{bc} \ , \end{equation}

\noindent the two Lorentz scalars become

\begin{equation} I_1 = \frac{1}{2}(E^2 - B^2) \hspace{2cm} 
I_2 = E\cdot B \ . \end{equation}

These scalars can now be used as building blocks to construct Lorentz 
invariant actions. In general we consider a Lagrangian density

\begin{equation} L = L(I_1, I_2) \ . \end{equation}

\noindent Two choices of special interest are

\begin{equation} L = I_1 \hspace{38mm} \mbox{(Maxwell)} 
\end{equation}

\begin{equation} L = 1 - \sqrt{1 - 2I_1 - I_2^2} \hspace{2cm} 
\mbox{(Born-Infeld)} \ . \end{equation}

\noindent The first one is familiar. The second is is the 
Lagrangian of the Born-Infeld theory, and can be rewritten in the 
elegant form

\begin{equation} L = 1 - \sqrt{- \det{(g_{{\alpha}{\beta}} + 
F_{{\alpha}{\beta}})}} \hspace{1cm} \mbox{(Born-Infeld)} \ . 
\end{equation}

We now consider the action

\begin{equation} S = \int \ L \ . \end{equation}

\noindent We wish to perform a Legendre transformation and find a 
phase space action of the form

\begin{equation} S = \int \ F_{ta}D_a - {\cal H}(I_1, I_2) \ , 
\end{equation}

\noindent where the invariants $I_1$ and $I_2$ are to be expressed 
as functions of the canonical variables $A_a$ and $D_a$ (and variation 
of the action with respect to $A_t$ gives Gauss' law as a constraint 
equation, as usual). For the Born-Infeld theory this was first done by 
Dirac \cite{Dirac}. The canonical momentum is the vector

\begin{equation} D_a = \frac{\partial L}{\partial I_1}E_a + 
\frac{\partial L}{\partial I_2}B_a \ . \end{equation}

\noindent For the two cases that we have singled out for attention 
we find that

\begin{equation} D_a = E_a \hspace{53mm} \mbox{(Maxwell)} 
\end{equation}

\begin{equation} D_a = \frac{1}{\sqrt{1 - 2I_1 - I_2^2}}
(E_a + I_2B_a) 
\hspace{2cm} \mbox{(Born-Infeld)} \ . \end{equation}

\noindent To actually express ${\cal H}$ as a function of $A_a$ 
and $D_a$ leads, in general, to a non-trivial calculation. For the 
Maxwell and Born-Infeld theories the required calculations show that

\begin{equation} {\cal H} = \frac{1}{2}(D^2 + B^2) \hspace{45mm} 
\mbox{(Maxwell)} \end{equation}

\begin{equation} {\cal H} = \sqrt{1 + D^2 + B^2 + D^2B^2 - 
(D\cdot B)^2} - 1 
\hspace{13mm} \mbox{(Born-Infeld)} \ . \end{equation}

When we give the McClain-Wu-Yu formulation of the Born-Infeld 
theory (in section 4) we will do so directly in the Hamiltonian 
formalism, without going through any Lagrangian preparations. It is 
therefore useful to know that the equation that shows that the 
Born-Infeld theory is Poincar\'{e} invariant 
is the current algebra relation

\begin{equation} \{ {\cal H}[N], {\cal H}[M]\} = \int \ 
(N\partial_aM - M\partial_aN){\epsilon}_{abc}D_bB_c \ , 
\end{equation}

\noindent where the notation ${\cal H}[N]$ denotes smearing 
with a test function. (It is not difficult to derive this relation 
provided that one sets about it 
in the right way, as we will in section 4.) On the right hand side 
we see the Poynting vector smeared with a particular test function 
formed from $N$ and $M$. For suitable choices of test functions the 
Poynting vector is the canonical generator of spatial translations 
and rotations. Since these are manifestly 
realized by the vector notation only the algebra of the Lorentz 
boosts needs to be checked explicitly; this algebra can be obtained 
by a suitable choice of the test functions in the relation that we 
just derived, which is why the 
latter is enough to ensure Poincar\'{e} invariance of the theory. 

We now turn to electromagnetic duality in these theories (following 
Deser and Teitelboim \cite{Deser}). It is well known that the 
Chern-Simons functional 

\begin{equation} {\omega} = \frac{1}{2}\int \ {\epsilon}_{abc}A_a
\partial_bA_c = \frac{1}{2} \int \ A_aB_a \end{equation}

\noindent (which is gauge invariant up to a surface term) 
generates the canonical transformation

\begin{equation} {\delta}A_a = 0 \hspace{1cm} {\delta}D_a = \{D_a, 
{\omega}\} = - B_a \ . \end{equation}

\noindent This is not, however, a symmetry of the action. Before 
we remedy this we rewrite the Chern-Simons functional in a 
non-local form. It is not difficult to show that

\begin{equation} {\omega} = \frac{1}{2} \int \ {\epsilon}_{abc}
B_a\frac{1}{\triangle}\partial_bB_c \ , \end{equation}

\noindent where ${\triangle}$ denotes the Laplace operator. Now, 
on the constraint surface of our electromagnetic theories the 
divergence of $D_a$ vanishes, and hence it can be expressed as 
the curl of a vector field, just as it is possible to do so for 
$B_a$ which is divergence free by definition. We can therefore 
introduce an "electric" Chern-Simons functional as well. Since 
$D_a$ is one of our canonical variables we prefer to use the 
non-local form of this functional; we choose to consider

\begin{equation} {\Omega} = \frac{1}{2} \int \ {\epsilon}_{abc}
(A_a\partial_bA_c - D_a\frac{1}{\triangle}\partial_bD_c) \ . 
\end{equation}

\noindent This functional generates the canonical transformations

\begin{equation} {\delta}A_a = - {\epsilon}_{abc}\frac{1}{\triangle}
\partial_bD_c \hspace{1cm} {\delta}B_a \approx D_a \hspace{1cm} 
{\delta}D_a = - B_a \ , \end{equation}

\noindent where the weak equality sign $\approx $ denotes equality 
modulo Gauss' law.

It is easy to show that

\begin{equation} \{{\Omega}, H_C\} = 0 \end{equation}

\noindent (where $H_C$ is the canonical Hamiltonian) for both 
the Maxwell and the Born-Infeld theory, hence ${\Omega}$ generates 
a symmetry of these theories. This symmetry is called 
electromagnetic duality. For a generic choice $L(I_1, I_2)$ of the 
Lagrangian density the requirement that ${\Omega}$ shall generate a 
symmetry leads to a differential equation for $L$ whose space 
of solutions has been studied by Gibbons and Rasheed \cite{Gibbons}. 
The duality of the Born-Infeld theory was first noted by 
Schr\"{o}dinger \cite{Erwin}. Note that the whole discussion rests 
on the constraint equation

\begin{equation} \partial \cdot D \approx 0 \ . \end{equation}

\noindent Hence we have no right to expect this symmetry to survive 
the introduction of sources or charged matter; there is no known 
analogue in Yang-Mills theory \cite{Deser}. 

There is another way in which to view electromagnetic duality. 
By definition of $B_a$ and $E_a$ we have the equations

\begin{equation} \partial \cdot B = 0 \hspace{2cm} \partial_tB_a 
= {\epsilon}_{abc}\partial_bE_c \ . \end{equation}

\noindent If we define 

\begin{equation} H_a \equiv \frac{{\delta}H_C}{{\delta}B_a} 
\end{equation}

\noindent we can write the remaining field equations in the form

\begin{equation} \partial \cdot D = 0 \hspace{2cm} \partial_tD_a 
= - {\epsilon}_{abc}\partial_bH_c \ . \end{equation}

\noindent Evidently the whole set of equations is left invariant 
by the duality transformation supplemented by

\begin{equation} {\delta}H_a \approx E_a \hspace{1cm} {\delta}E_a 
\approx - H_a \ . \end{equation} 

\noindent But these equations follow from the vanishing of 
$\{ {\Omega}, H_C\}$. This is in fact a common way in which to 
view electromagnetic duality.

Our aim, which is to express the canonical generator of 
electromagnetic duality in local form, will be reached in the next 
section; before we come to it it will be useful to give the 
two-potentials formulation of the Maxwell theory \cite{Zwanziger}. 
(The two-potentials formulation of the Born-Infeld theory can be 
easily derived from the equations given in section 4, and will 
be omitted here.) The introduction of two vector potentials to 
describe electromagnetism in fact a rather natural thing to try, 
since the obstruction to extending duality to a theory with sources 
has to do with the fact that the electric field is no longer 
divergence free and consequently can not be 
written as the curl of a vector potential. On the other hand 
(taking a four dimensional point of view) it is true that an 
arbitrary two-form can always be written as the sum of the "curls" 
of two independent vector potentials \cite{Plebanski}. A direct 
way to arrive at a model with two vector potentials is to 
consider the action for an anti-symmetric tensor in six space-time 
dimensions, and decompose it into four dimensional fields

\begin{equation} (A_{{\alpha}{\beta}}, A_{{\alpha}4}, 
A_{{\beta}5}, A_{45}) \equiv \frac{1}{2\sqrt{2}}({\phi}_{ab}, 
A_{a1}, A_{a2}, {\phi}) \ . \end{equation}

\noindent If we perform dimensional reduction to four dimensions 
of the standard action for the anti-symmetric tensor in six 
dimensions, and then throw away the terms involving 
${\phi}_{ab}$ and ${\phi}$, we arrive at

\begin{equation} S = \frac{1}{2} \int \ F_{ta1}D_{a1} + 
F_{ta2}D_{a2} - \frac{1}{2}(E_{a1}E_{a1} + D_{a2}D_{a2} + 
B_{a1}B_{a1} + B_{a2}B_{a2}) \ . \end{equation}

\noindent This action has an obvious symmetry generated by the 
weakly gauge invariant functional

\begin{equation} \tilde{\Omega} = - \frac{1}{2}\int 
A_{a1}D_{a2} - A_{a2}D_{a1} \ . \end{equation}

\noindent This is a perfectly local functional, whose effect 
is to rotate the two vector potentials into each other.

We can recover the Maxwell theory by constraining the two-potentials 
theory suitably. Again this has a natural six dimensional 
interpretation; what one has to do is to restrict the theory to 
anti-symmetric tensors whose field strengths are self-dual under 
six dimensional Hodge duality \cite{Marcus}. (Such objects are 
called chiral p-forms.) After dimensional reduction, Hodge 
self-duality of the anti-symmetric tensor implies that 

\begin{equation} E_{a1} = B_{a2} \hspace{2cm} E_{a2} = - B_{a1} 
\ . \end{equation}

\noindent In the two-potentials theory this is equivalent to 
the constraints

\begin{equation} {\Phi}_{a1} = D_{a1} - B_{a2} \approx 0 
\hspace{1cm} 
{\Phi}_{a2} = D_{a2} + B_{a1} \approx 0 \ . \end{equation} 

\noindent These constraints can be implemented by means of 
Lagrange multiplier terms added to the action, if one desires 
to do so. Note that 

\begin{equation} \{{\Phi}_{a1}, H\} = - 2{\epsilon}_{abc}
\partial_b{\Phi}_{c2} \approx 0 \hspace{1cm} 
\{{\Phi}_{a2}, H\} = 2{\epsilon}_{abc}\partial_b{\Phi}_{c1} 
\approx 0 \ . \end{equation}

\noindent Preservation of the constraints under time evolution 
then forces the transverse parts of their corresponding Lagrange 
multipliers to vanish in a solution. This means that the 
equations of motion are unchanged by 
this manoeuvre; the constraints select a subspace of the space 
of solutions, but leave the solutions themselves unchanged. The 
non-zero part of the constraint algebra is

\begin{equation} \{{\Phi}_{a1}(x), {\Phi}_{b2}(y)\} = - 
4{\epsilon}_{acb}\partial_c{\delta}(x,y) \ . \end{equation}

\noindent We see that the constraints are a mixture of first and 
second class constraints. They contain a first class component 
because they imply Gauss' law; we can get rid of this complication 
by setting the longitudinal parts of the vector potentials to zero, 
so that Gauss' law holds strongly. Assume that this has been done. 
The remaining constraints are purely second 
class, and can be solved by

\begin{equation} E^T_{a2} = - B_{a1} \hspace{1cm} A^T_{a2} = 
- {\epsilon}_{abc}\frac{1}{\triangle}B_{c2} = - {\epsilon}_{abc}
\frac{1}{\triangle}D^T_{c1} \ , \end{equation}

\noindent where $T$ denotes the transverse part. When we insert 
this result in the expression for the generator $\tilde{\Omega}$ 
the latter becomes non-local. This non-locality is 
not really due to the Coulomb gauge --- it is in fact impossible 
to solve the second class component of the constraints in a local 
manner. 

When we insert the solution of the constraints into the action 
for the two-potentials theory we obtain

\begin{equation} S = \int \ \dot{A}^T_{a1}D^T_{a1} - \frac{1}{2}
(D^T_{a1}D^T_{a1} + B_{a1}B_{a1}) \ . \end{equation}

\noindent But this is precisely the action for Maxwell's theory 
in the Coulomb gauge. Hence the constrained version of the 
two-potentials theory 
is equivalent to Maxwell's. Unsurprisingly, we also find that 

\begin{equation} \tilde{\Omega} = {\Omega} \ . \end{equation}

\noindent Thus the local symmetry generator $\tilde{\Omega}$ 
becomes the non-local generator of electromagnetic duality after 
constraining the theory.

The advantage of the two-potentials formulation is that it makes 
electromagnetic duality manifest - the latter is now generated by 
a local functional. The way in which this advantage was gained 
may strike the reader as a fake. Indeed it is a fake in a sense, 
because second class constraints implying a non-local symplectic 
structure were included in the bargain, and we do not have a 
consistent Hamiltonian system until these have been solved for. 
But in another sense it is not, because the two-potentials 
formulation is the germ of the fully local as well as 
manifestly duality invariant formulation to be reviewed in the 
next section.

\vspace{1cm}

\noindent {\bf 3. LOCAL DUALITY IN THE MAXWELL CASE.}

\vspace{5mm}

\noindent As we have seen the locality of the two-potentials 
formulation of electrodynamics is spoilt by the presence of 
second class constraints. Now the McClain-Wu-Yu formulation 
was originally obtained by following Batalin's and Fradkin's 
algorithm \cite{Batalin} for replacing second class 
constraints with gauge symmetries and new degrees of freedom. 
It was first applied to chiral bosons in $1+1$ dimensions 
\cite{McClain}, then to electrodynamics (by Martin and 
Restuccia \cite{Martin}) and finally to chiral p-forms in 
twice odd dimensions \cite{Henneaux}, where it solves the 
problem \cite{Marcus} of giving a manifestly covariant 
formulation for such fields. It is interesting to observe that 
the McClain-Wu-Yu formulation emerges naturally from string 
field theory, where the infinite set of gauge fields arises 
because of the presence of a bosonic ghost zero mode in the 
Ramond-Ramond sector \cite{Berkovits}. (For completeness we 
mention that there exists an alternative approach to manifestly 
covariant chiral p-forms \cite
{Sorokin}, but this will play no role here.)

We devote this section to a brief but careful review of these 
matters for the Maxwell theory. The canonical variables are 
an infinite set of pairs of vector potentials and their conjugate 
momenta, indexed by $(n)$ and $i$, where the index $i$ takes 
the values $1$ and $2$. The phase space action is

\begin{equation} S = \frac{1}{2} \sum_{n=0}^{\infty }\int \ 
\dot{A}_{ai}^{(n)}D_{ai}^{(n)} - \frac{(-1)^n}{2}(E_{ai}^{(n)}
E_{ai}^{(n)} + B_{ai}^{(n)}B_{ai}^{(n)}) - 
{\Lambda}_{ai}^{(n+1)}{\Psi}_{ai}^{(n+1)} - {\Lambda}_i^{(n)}
{\cal G}_i^{(n)} \ , \end{equation}

\noindent where summation over $i$ (and the vector indices) is 
understood. This action can be derived from the manifestly 
covariant Lagrangian for a chiral two-form in six dimensions 
\cite{Kleppe} by dimensional reduction followed by a Legendre 
transformation. Varying the action with respect to the Lagrange 
multipliers gives rise to an 
infinite set of constraints, which by definition are

\begin{equation} {\cal G}_i^{(n)} \equiv \partial_aD_{ai}^{(n)} 
\approx 0 \end{equation}

\begin{equation} {\Psi}_{ai}^{(n+1)} \equiv {\Pi}_{ai}^{-(n)} + 
{\Pi}_{ai}^{+(n+1)} \approx 0 \ , \end{equation}

\noindent where we use the useful further definitions

\begin{equation} {\Pi}_{ai}^{\pm (n)} \equiv E_{ai}^{(n)} \pm 
{\epsilon}_{ij}B_{aj}^{(n)} \ . \end{equation}

\noindent It is straightforward to check that all the 
constraints are first class and that 

\begin{equation} \{{\Psi}_{ai}^{(n+1)}, H_C\} = (-1)^{n+1}
{\epsilon}_{ij}{\epsilon}_{abc}\partial_b{\Psi}_{cj}^{(n+1)} 
\approx 0 \ , \end{equation}

\noindent where $H_C$ is the canonical Hamiltonian. It follows 
that this may be a consistent Hamiltonian system. However, before 
we can conclude that this is indeed the case we must give a more 
careful definition of its phase space, to ensure that all 
the relevant infinite sums converge \cite{Kleppe}.

By the way we observe that if one attempts to treat a non-abelian 
gauge theory in the same way one finds that the constraint algebra 
does not close --- the bracket between Gauss' law and the "new" 
constraints will not behave itself. If there is a generalization 
of all this to Yang-Mills theory, it is a quite non-trivial one.

Returning to electrodynamics we require that its canonical 
Hamiltonian exists, which is ensured if for any set of fields 
there exists an $N$ such that 

\begin{equation} n > N \hspace{1cm} \Rightarrow \hspace{1cm} 
|{\Pi}_{ai}^{\pm (n)}| \leq \frac{f}{n} \ , \end{equation}

\noindent where $f(x)$ is some square integrable function. (A 
convenient choice is $f(x) = 0$ \cite{Nathan}). We will only 
allow initial data obeying this condition, and we then have to 
show that it is preserved under time evolution and gauge 
transformations. This will be so only if we 
restrict the allowed gauge transformations in a suitable manner. 
The precise statement is that the formal expression

\begin{equation} \sum_{n=0}^{\infty }{\Psi}^{(n+1)}_{ai}
[{\Lambda}_{ai}^{(n+1)}] \equiv \sum_{n=0}^{\infty }\int 
\ {\Lambda}_{ai}^{(n+1)}{\Psi}^{(n+1)}_{ai} 
\end{equation}

\noindent is a generator of allowed gauge transformations, and 
hence a first 
class constraint, only if there exists an $N$ such that

\begin{equation} n > N \hspace{1cm} \Rightarrow \hspace{1cm} 
|{\Lambda}_{ai}^{\pm (n+1)}| \sim \frac{1}{n} \ . \end{equation}

\noindent When this requirement is kept firmly in mind various 
"traps" are avoided; the McClain-Wu-Yu formulation is indeed a 
consistent one.

In particular we can now show that an allowed set of gauge 
choices is

\begin{equation} A^{L(n+1)}_{ai} = 0 \hspace{1cm} \Rightarrow 
\hspace{1cm} E^{L(n+1)}_{ai} = 0  \end{equation}

\begin{equation} {\Pi}^{+(n+1)}_{ai} = 0 \hspace{1cm} 
\Rightarrow \hspace{1cm} 
{\Pi}^{-(n)}_{ai} = 0 \ , \end{equation}

\noindent where $L$ denotes the longitudinal part. When these 
conditions are inserted into the phase space action we recover 
the two-potentials formulation of the Maxwell theory, including 
its second class constraints, and the 
equivalence to the original Maxwell theory follows.

Before gauge fixing, the generator of electromagnetic duality 
transformations is the local functional

\begin{equation} \tilde{\Omega} = - \frac{1}{2} \sum_{n=0}^
{\infty } \int \ 
{\epsilon}_{ij}A_{ai}^{(n)}D_{aj}^{(n)} \ . \end{equation}

\noindent (Which exists.) One can check that

\begin{equation} \{ {\Psi}_{ai}^{(n+1)}, \tilde{\Omega}\} = 
\frac{1}{2}
{\epsilon}_{ij}{\Psi}_{aj}^{(n+1)} \approx 0 \end{equation}

\begin{equation} \{ \tilde{\Omega}, H_C\} = 0 \ . 
\end{equation}

\noindent Hence $\tilde{\Omega}$ is a weakly gauge invariant 
symmetry generator, and it easy to see that it reduces to the 
non-local symmetry generator ${\Omega}$ 
when all the gauges have been fixed. We can therefore conclude 
that the McClain-Wu-Yu formulation indeed leads to a 
formulation of the Maxwell theory in which electromagnetic 
duality is realized as a fully local symmetry.

\vspace{1cm}

\noindent {\bf 4. LOCAL DUALITY IN THE BORN-INFELD CASE.}

\vspace{5mm}

\noindent Finally we are prepared to deal with the subject of 
our paper, that is manifest duality of the Born-Infeld theory. 
This theory differs from Maxwell's 
only in the choice of the Hamiltonian, so we use the same phase 
space --- including the definition of allowed gauge transformations 
and the generator $\tilde{\Omega}$ of duality transformations --- 
as was introduced in the previous section. It is interesting to ask 
whether one can derive the Born-Infeld Hamiltonian by dimensional 
reduction from a suitably constrained action for a two-form in six 
dimensions, but it is not easy to construct such an action 
\cite{Perry} --- the analogue of the invariant $I_2$ does not exist 
in twice odd dimensions. For this reason we will instead 
make a reasonable guess for the Hamiltonian, verify that it leads 
to a consistent and Poincar\'{e} invariant theory, and check that 
the ordinary Born-Infeld 
theory can be derived from it by gauge fixing.

The obvious guess for the Hamiltonian density is

\begin{equation} {\cal H} = \sqrt{1 + 2{\cal H}^M + 
{\cal H}^M_a{\cal H}_a^M} - 1 \ , \end{equation}

\noindent where 

\begin{equation} {\cal H}^M = \sum_{n=0}^{\infty } 
\frac{(-1)^n}{4}(E_{ai}^{(n)}E_{ai}^{(n)} + B_{ai}^{(n)}
B_{ai}^{(n)}) \ , \end{equation}

\begin{equation} {\cal H}_a^M = \sum_{n=0}^{\infty } 
\frac{1}{2}{\epsilon}_{abc}E^{(n)}_{bi}B_{ci}^{(n)} \ , 
\end{equation}

\noindent and the superscript $M$ may stand for Maxwell, 
McClain-Wu-Yu, or Martin and Restuccia according to preferences. 
Given our definition of the phase space these objects, and 
hence the Hamiltonian, clearly exist.

To see whether time evolution preserves the phase space, and to 
see whether Poincar\'{e} invariance is present, it is convenient 
to define the functional derivatives

\begin{equation} \frac{{\delta}{\cal H}[N]}{{\delta}
A_{ai}^{(n)}} = \frac{1}{2}{\epsilon}_{abc}\partial_b\left(
\frac{N}{{\cal H} + 1}((-1)^nB_{ci}^{(n)} - {\epsilon}_{cde}
D_{di}^{(n)}{\cal H}^M_e)\right) \end{equation}

\begin{equation} \frac{{\delta}{\cal H}[N]}{{\delta}
D_{ai}^{(n)}} = \frac{1}{2}\frac{N}{{\cal H} + 1}((-1)^n
D_{ai}^{(n)} + {\epsilon}_{abc}
B_{bi}^{(n)}{\cal H}^M_c) \ . \end{equation}

\noindent We can now see by inspection that the phase space 
is preserved by time evolution. Poincar\'{e} invariance is a 
little bit more subtle. Given the functional derivatives it is 
straightforward to write down the crucial 
current algebra relation

\begin{eqnarray} \{ {\cal H}[N], {\cal H}[M]\} = {\cal H}_a^M
[N\partial_aM - M\partial_a N] - \hspace{3cm} \nonumber \\
\ \\ 
- \sum_{n=0}^{\infty } \int \ (N\partial_aM - M\partial_a N)
(-1)^n(D_{ai}^{(n)}D_{bi}^{(n)} + B_{ai}^{(n)}B_{bi}^{(n)})
{\cal H}^M_b \ . \nonumber \end{eqnarray}

\noindent The last term on the right hand side ought not to be 
there. However, we can show that

\begin{eqnarray} \sum_{n=0}^{\infty } (-1)^n(D_{ai}^{(n)}
D_{bi}^{(n)} + B_{ai}^{(n)}B_{bi}^{(n)}) = \frac{1}{2}{\Pi}^
{+(0)}_{ai}{\Pi}^{+(0)}_{bi} + \hspace{2cm} \nonumber \\
\ \\
 + \frac{1}{4}\sum_{n=0}^{\infty } (-)^n\left( (
{\Pi}^{-(n)}_{ai} - {\Pi}^{+(n+1)}_{ai}){\Psi}_{bi}^{(n+1)} 
+ {\Psi}_{ai}^{(n+1)}({\Pi}^{-(n)}_{bi} - {\Pi}^{+(n+1)}_{bi})
\right) \nonumber . \end{eqnarray}

\noindent By inspection of the terms that multiply the 
constraints on the right hand side we see that their behaviour 
for increasing $n$ is that of the parameters for an allowed gauge 
transformation. Therefore --- but only 
therefore --- we may conclude that

\begin{equation} \sum_{n=0}^{\infty } (-1)^n(D_{ai}^{(n)}
D_{bi}^{(n)} + B_{ai}^{(n)}B_{bi}^{(n)}) \approx \frac{1}{2}
{\Pi}^{+(0)}_{ai}{\Pi}^{+(0)}_{bi} \ . \end{equation}

\noindent Proceeding in a similar way, we see that

\begin{equation} {\cal H}^M_a \approx - \frac{1}{8}{\epsilon}
_{abc}{\epsilon}_{ij}{\Pi}^{+(0)}_{bi}{\Pi}^{+(0)}_{cj} \ , 
\end{equation}

\noindent and finally that

\begin{equation} \{ {\cal H}[N], {\cal H}[M]\} \approx 
{\cal H}_a^M[N\partial_aM - M\partial_a N] \ . \end{equation}

\noindent The conclusion that the Hamiltonian that we have 
defined indeed leads to a Poincar\'{e} invariant theory 
follows.

The rest of the required arguments follow quickly. It is clear 
that the duality generator $\tilde{\Omega}$ of the previous 
section generates a symmetry also of this theory. That the 
theory is equivalent to the Born-Infeld theory is evident when 
we fix the gauges, again as in the previous 
section. Having done so we find that

\begin{equation} {\Pi}^{+(0)}_{a1} = 2D^{(0)}_{a1} \hspace{1cm} 
{\Pi}^{+(0)}_{a2} = 2B^{(0)}_{a2} \ . \end{equation}

\noindent The unconstrained pair $(A_{a1}^{(0)}, 
D_{b1}^{(0)})$ obey the canonical Dirac brackets of the 
Maxwell theory. Moreover we have already shown that

\begin{equation} {\cal H}^M \approx \frac{1}{8}{\Pi}^{+(0)}
_{ai}{\Pi}^{+(0)}_{ai} \approx \frac{1}{2}(D^{(0)}_{a1}D^{(0)}
_{a1} + B^{(0)}_{a1}B^{(0)}_{a1}) \end{equation}

\begin{equation} {\cal H}^M_a \approx {\epsilon}_{abc}
D^{(0)}_{b1}B^{(0)}_{c1} \ . \end{equation}

\noindent Hence our Hamiltonian agrees with the Born-Infeld 
Hamiltonian on the constraint surface, and the two theories 
are indeed equivalent.

\vspace{1cm}

\noindent {\bf 5. CONCLUSION.}

\vspace{5mm}

\noindent Through the introduction of an infinite set of gauge 
fields it is possible to formulate the Maxwell theory in such a 
way that electromagnetic duality is generated by a local 
functional. We showed --- paying attention to the precise 
definition of the phase space --- that the same formulation works 
for the non-linear Born-Infeld theory. 

We did not treat sources (but see the papers by Berkovits 
\cite{Nathan} for this). Whether there are other and perhaps 
more interesting dualities that can be treated in a similar way 
is an open question, but this is an issue that may 
be worth thinking about.

\end{document}